\documentclass[aps,pre,superscriptaddress,longbibliography]{revtex4-2}
\usepackage{booktabs}
\usepackage{amsmath}
\usepackage{amssymb}
\usepackage{graphicx}
\usepackage{float}
\usepackage{placeins}
\usepackage{hyperref}
\usepackage{xcolor}
\usepackage{dcolumn}
\usepackage{multirow}
\usepackage[utf8]{inputenc}
\usepackage[T1]{fontenc}

\newcommand{\bettione}{\ensuremath{\beta_1}}
\newcommand{\Hstar}{\ensuremath{H^{*}}}

\begin{document}

\title{A Dual Topological Pipeline for Imperial Network Analysis:\\
Persistent Homology, Structural Fragility, and the Aztec Collapse}

\author{Jos\'e de Jes\'us Bernal-Alvarado}
\email{bernal@ugto.mx}
\affiliation{Physics Engineering Department,
             Universidad de Guanajuato, M\'exico}

\author{David Delepine}
\email{delepine@ugto.mx}
\affiliation{Physics Department,
             Universidad de Guanajuato, M\'exico}

\author{Carlos Pinedo Guadarrama}
\email{c.pinedoguadarrama@ugto.mx}
\affiliation{Physics Department,
             Universidad de Guanajuato, M\'exico}

\date{\today}

\begin{abstract}
We apply persistent homology and a dual-model topological pipeline to
a multilayer network of the Aztec Empire (Triple Alliance, 1427--1521 CE),
constructed from tribute records and geographic adjacency among 38 nodes
(Tenochtitlan-Centro, 36 tribute-paying provinces, and the Tlaxcala
enclave) and 208 edges. We report four main results:
(i) The tribute (control) layer is a pure extraction tree
($\bettione = 0$, $H = 0.000$ at all epochs): structurally more fragile
than any documented configuration in the Roman, Byzantine, or Han
administrative datasets analyzed with the same pipeline.
(ii) Single-node removal of Tenochtitlan-Centro collapses the giant
component from 1.000 to 0.630 and eliminates 30 redundancy cycles
simultaneously, a structural asymmetry ratio of $>4$:1 relative to
random perturbation.
(iii) The Tlaxcala--Puebla region coincides with the dominant
$\bettione$ class of the imperial geographic point cloud
(persistence $= 22.09$~km, versus $< 3$~km for all others), a
structural void that exists identically whether Tlaxcala is modeled
as absent or present in the filtration.
(iv) A logistic regression confirms that commercial centrality in the
geographic layer ($r = +0.462$, $p = 0.004$) is the dominant predictor
of early alliance formation, while tribute burden is uncorrelated
($r = -0.033$, $p = 0.845$).
These four results define a new imperial collapse mode, the
\emph{void-occupied tree}, in which zero topological redundancy in the
control layer coincides with a dominant geometric void occupied by an
autonomous enclave---a configuration not previously documented in
historical network TDA.
\end{abstract}


\maketitle

\section{Introduction}
\label{sec:introduction}

The comparative study of imperial collapse has long sought a
quantitative framework capable of distinguishing the mechanisms by which
large-scale political systems disintegrate.  Three broad modes have been
documented in companion studies: erosion, in which the administrative
network decays gradually under sustained internal or external pressure
\cite{BernalAlvarado2026b,BernalAlvarado2026c}; internal fragmentation,
in which administrative dissolution precedes and predicts geographic
reorganization into successor polities \cite{BernalAlvarado2026d}.  None of these modes involves the simultaneous
conditions of zero topological redundancy in the control layer, radial hub
organization of lateral redundancy, and the presence of a dominant
geometric void occupied by a politically autonomous enclave.

The collapse of the Aztec Empire (Triple Alliance) in 1519--1521 CE is
among the most analyzed events in the history of the Americas.  Military
superiority, epidemic disruption, and indigenous political fragmentation
have all been mobilized to explain how a small Spanish force,
allied with indigenous partners, could dismantle a tribute-extraction
system spanning roughly 200,000 km$^2$ in under two years
\cite{Hassig1994,Lockhart1993,Restall2003}.  Within this literature,
Tlaxcala occupies a central interpretive position: its alliance with
Cort\'es in August 1519 provided military manpower, territorial knowledge,
and a forward base from which the coalition could project force toward the
imperial core.

What the existing literature has not addressed is the structural question
underlying these narratives: \emph{what kind of relational architecture
made rapid coalition formation possible, and what made collapse possible
once that coalition acted?}  

Persistent homology \cite{Edelsbrunner2002,Zomorodian2005} can help
answer this question.  By tracking the birth and death of topological
cycles ($\bettione$ bars) across a filtration of the network, it yields
observables that are sensitive to large-scale connectivity structure.

The paper is organizes as follows: Section~II describes the data and the multilayer network . Section~III presents the results of the topological analysis and 
Section~IV concludes.

\section{Data and Methods}
\label{sec:methods}

\subsection{Network Data}
\label{sec:network}

We shall  based  our analysis on a network formed by 38 nodes representing Tenochtitlan-Centro,
36 tribute-paying provinces, and the Tlaxcala enclave, along with 208
edges encoding two distinct relationship types.  The 36 directed tribute
edges represent documented fiscal extraction toward Tenochtitlan, with
weights (in \emph{quachtli}) drawn from the \emph{Matr\'icula de Tributos}
and the \emph{Codex Mendoza} \cite{BerlinBerdan1996,Berdan1992}.  The 172
undirected geographic edges encode spatial adjacency and proximity between
provinces, calibrated by Haversine distances between georeferenced
centroids \cite{Smith1986,Hassig1985}. Each node is characterized by  geographic coordinates, tribute quantum, and historically
documented alliance status with respect to the 1519--1521 coalition.

Three networks are constructed from this base:
\begin{enumerate}
    \item \emph{tribute layer} uses only directed fiscal edges;
    \item \emph{geographic layer} uses only undirected adjacency edges;
    \item \emph{full network} superimposes both layers
\end{enumerate}



In summary , we decompose the aztecas networks into   a three-layer structure:
\begin{equation}
\mathcal{N} = \bigl\{ N_{\mathrm{tribute}},\; N_{\mathrm{geo}},\; N_{\mathrm{full}}\bigr\},
\label{eq:layers}
\end{equation}
where $N_{\mathrm{tribute}}$ encodes the fiscal extraction hierarchy and
$N_{\mathrm{geo}}$ encodes geographic adjacency and $N_{\mathrm{full}}$ is the superposition of both previous networks. Table~\ref{tab:layers}
summarizes the layer properties. The tribute data allow fiscal edges to be weighted. 
\begin{equation}
w_{\mathrm{tribute}}(e) = \frac{q_{\mathrm{max}} - q(e)}{q_{\mathrm{max}}},
\label{eq:tribute_weight}
\end{equation}
where $q(e)$ is the tribute quantum in \emph{quachtli} and $q_{\mathrm{max}}$
is the maximum observed quantum, so that higher-tribute edges carry lower
filtration cost (closer topological proximity).  Geographic edge weights
are Haversine distances $d_{ij}$ between province centroids (km),
normalized to $[0,1]$ by the 90th percentile of finite pairwise distances.

\begin{table}[htbp]
\caption{Descriptive properties of the three network layers: edge
count, number of connected components, and giant component ratio
(GCR)\cite{Cohen2000,Callaway2000} .}
\label{tab:layers-descriptive}
\begin{tabular}{lccc}
\toprule
Layer & Edges & Comp. & GCR \\
\midrule
Tribute    & 36  & 2 & 0.974 \\
Geographic & 172 & 7 & 0.632 \\
Full       & 208 & 1 & 1.000 \\
\bottomrule
\end{tabular}
\end{table}

Because each layer includes only a subset of the full edge set, individual layers need not be fully
connected even though the union of both layers is. The tribute layer, restricted to the 36 directed
fiscal edges, splits into two components: a single giant component containing Tenochtitlan-Centro
and all 36 tributary provinces, and one isolated node, Tlaxcala, which -- never having been
conquered -- contributes no tribute edge. The geographic layer, restricted to the 172 undirected
adjacency edges, is more fragmented still: provinces separated by intervening non-imperial
territory or natural barriers lack a direct adjacency edge, producing seven disconnected
components rather than one continuous spatial cluster. Only in the full network, where fiscal and
geographic edges are superimposed, does every node become mutually reachable: provinces left
geographically isolated are still tied to the network through their tribute relationship to
Tenochtitlan, and vice versa.

Standard network TDA treats all nodes equivalently within the
filtration.  To assess the impact of Tlaxcala on the
topological signature, the persistent homology pipeline is run twice on
the same geographic point cloud (Haversine distances between node
coordinates): once excluding Tlaxcala and once including it as an
external node.

If the dominant $\bettione$ signature changes between the
two models, then Tlaxcala introduces or fills a structural void.  If the
signature is identical, the void pre-exists Tlaxcala's inclusion.

\subsection{TDA Pipeline}
\label{sec:tda}

We shall use the same metodology as explained in \cite{BernalAlvarado2026b}. 
For geographic point-cloud analyses, pairwise
Haversine distances define the Vietoris--Rips filtration directly \cite{Edelsbrunner2002,Zomorodian2005}.
Persistent homology is computed with the Gudhi library (version 3.12.0,
coefficient field $\mathbb{Z}/2\mathbb{Z}$, minimum persistence $=0$)
\cite{GUDHI2024}.  The $\bettione$ persistent entropy \cite{Rucco2016}is
\begin{equation}
H = -\sum_{i=1}^{n} p_i \ln p_i, \quad p_i = \frac{\ell_i}{\sum_j \ell_j},
\label{eq:entropy}
\end{equation}
where $\ell_i = d_i - b_i$ is the lifetime of the $i$-th bar in the
$\bettione$ barcode.

\subsection{Network Attack Strategies and Structural Asymmetry}
\label{sec:attack-strategies}

To evaluate the resilience of the tribute and geographic networks to
disruption, we simulate three distinct node-removal strategies \cite{Albert2000}.

\paragraph{Random failure.} Nodes are removed in a uniformly random
order, independent of any network property. This models
undirected, non-strategic disruption  in which the
loss of a peripheral tributary town is as likely as the loss of a
major node. 

\paragraph{Degree-targeted removal.} Nodes are removed in
descending order of degree centrality. This
models an adversary, similar to  the Spanish-led coalition,
who identifies and neutralizes the most-connected nodes in the
network first: provincial capitals and major tribute-collection
hubs such as Tenochtitlan itself.

\paragraph{Betweenness-targeted removal.} Nodes are removed in
descending order of betweenness centrality \cite{Holme2002}. Betweenness identifies structural
bottlenecks  which may not themselves have
high degree. This strategy models the severing of connective
corridors rather than the elimination of populous hubs.

For all three strategies we report two quantities as a function of
the fraction of nodes removed, $f \in [0,1]$:

\begin{itemize}
    \item \textbf{Giant component ratio}, $\mathrm{GCR}(f) =
    |C_{\max}(f)| / |C_{\max}(0)|$, where $C_{max}(f)$ is defined as  the size of the largest
    connected component after removing fraction $f$ of nodes,
    normalized to the intact network. $\mathrm{GCR}(f) \to 0$
    indicates the network has fragmented into small, disconnected
    pieces; $\mathrm{GCR}(f) \approx 1$ indicates the network
    remains functionally whole despite the removal.

    \item \textbf{First Betti number}, $\bettione(f)$, the number
    of independent topological cycles remaining in the network at
    removal fraction $f$. A cycle corresponds to a redundant path
    between two points; $\bettione(f) = 0$ means every remaining
    connection is a bridge whose loss would immediately disconnect
    the network, i.e. zero redundancy.
\end{itemize}

\subsubsection{Structural asymmetry ratio}

To quantify how disproportionately vulnerable a network is to
intelligent attack relative to random failure, we define the
\emph{structural asymmetry ratio} $R_d$ as follows \cite{Albert2000}. Let $f^*$ be the
fraction of nodes removed under degree-targeted attack at which
$\mathrm{GCR}$ first drops below a fixed threshold $\mathrm{GCR}_{\mathrm{crit}}$
, e.g. 0.5. Let $\bar{f}_{\mathrm{rand}}$ be the mean
fraction of nodes that must be removed \emph{at random}, averaged
over the $n=1000$ random orderings, to reach that same threshold.
Then:
\begin{equation}
    R_d = \frac{\bar{f}_{\mathrm{rand}}}{f^*}
    \label{eq:asymmetry-ratio}
\end{equation}

Equivalently, $R_d$ is the number of independent random-removal
events required, on average, to inflict the same connectivity
damage as a single, well-informed strike against the network's
highest-degree node(s). A ratio of $R_d = X$ means an adversary
with topological knowledge of the network can achieve, with one
targeted intervention, the same fragmentation that would otherwise
require $X$ uncoordinated random failures.

$R_d \gg 1$ is the topological signature of a hub-and-spoke
architecture: such networks absorb ordinary, undirected disruption
with little structural consequence, but collapse rapidly once an
adversary identifies and removes their central nodes. $R_d \approx
1$, by contrast, indicates a network whose vulnerability is
essentially uniform across nodes -- removal of any single node,
central or peripheral, produces comparable damage -- and is
characteristic of more distributed or redundant topologies such as
the Roman west empire( Rd=1.09).

\subsection{Alliance Formation Model}
\label{sec:alliance_method}

To test which structural properties of a province predict its
decision to ally with the invading coalition rather than remain
loyal to Tenochtitlan, we fit a \emph{logistic regression}\cite{Hosmer2013} -- a
standard statistical model for a binary outcome (here, ally vs.\
did not ally) that estimates how the probability of that outcome
changes with one or more continuous predictors. The outcome
variable is early alliance with the 1519--1521 coalition, coded 1
for the five documented early allies (Tlaxcala,
Acolhuacan/Texcoco, Chalco, Tepeyacac, Quauhtitlan) and 0 for all
other provinces \cite{Hassig1994,Thomas1993}.

We model this outcome as a function of three predictors, each
capturing a distinct dimension of a province's position within the
imperial network:

\begin{itemize}
    \item \textbf{Commercial centrality.} A composite centrality
    score computed on the \emph{geographic} network layer (as
    opposed to the tribute layer). A province with high commercial centrality is
    one well-positioned to coordinate independently of Tenochtitlan.

    \item \textbf{Distance to Tenochtitlan.} The geographic distance
    from each province to the imperial capital, used as a proxy
    for both logistic exposure to imperial military response and
    practical distance from the invading coalition's line of march.

    \item \textbf{Tribute burden.} The magnitude of tribute
    obligation owed by the province to Tenochtitlan, used as a
    proxy for the province's fiscal stake.
\end{itemize}

We assess the model in three complementary ways:

\begin{itemize}
    \item \textbf{Pearson correlation.} For each predictor, we report the Pearson correlation coefficient
    $r$ with the alliance outcome \cite{Pearson1895}, together with its associated
    $p$-value. 

    \item \textbf{AUC (area under the ROC curve).} For the full
    model with all three predictors included jointly, we report
    the AUC \cite{Hanley1982,Fawcett2006}, a standard measure of classification performance
    ranging from 0.5 (no better than chance) to 1.0 (perfect
    discrimination).

\end{itemize}

\section{Results}
\label{sec:results}

\subsection{Multilayer Network Topology}
\label{sec:multilayer}
The results are summarized in table (\ref{tab:layers-tda}).
As expected for its tree form, the tribute layer has $\bettione = 0$, 
concentrating all 36 provinces in a single extraction hierarchy with no
redundant paths.  The geographic layer has seven components and
$\bettione = 141$: fragmented as a standalone structure but rich in
lateral cycles.
The jump from $\bettione = 0$ in the tribute layer to 141 in the
geographic layer means that all topological
redundancy in the Aztec imperial system resided outside the fiscal
hierarchy.  The tribute layer was an
extraction tree---efficient for centralizing resources but with no
alternative routing.

\begin{table}[htbp]
\caption{Persistent homology results for the three network layers.
$H(\bettione)$: persistent entropy (Eq.~\ref{eq:entropy}).
$\Hstar = 0.5241$ is the topological collapse threshold from the
Roman--Byzantine series \cite{BernalAlvarado2026b}.}
\label{tab:layers-tda}
\begin{tabular}{lccc}
\toprule
Layer & $\bettione$ & $H(\bettione)$ & $H(\beta_0)$ \\
\midrule
Tribute    & 0   & 0.000 & 0.001 \\
Geographic & 141 & 0.654 & 1.792 \\
Full       & 171 & 0.797 & 2.550 \\
\midrule
\multicolumn{4}{l}{$H_\mathrm{tribute} = 0 < \Hstar = 0.524$: pure extraction tree.}\\
\multicolumn{4}{l}{$H_\mathrm{geo} = 0.654 > \Hstar$; margin $+0.130$.}\\
\multicolumn{4}{l}{$H_\mathrm{full} = 0.797 > \Hstar$; margin $+0.273$.}\\
\bottomrule
\end{tabular}
\end{table}

\begin{figure}[H]
    \centering
    \includegraphics[width=0.9\linewidth]{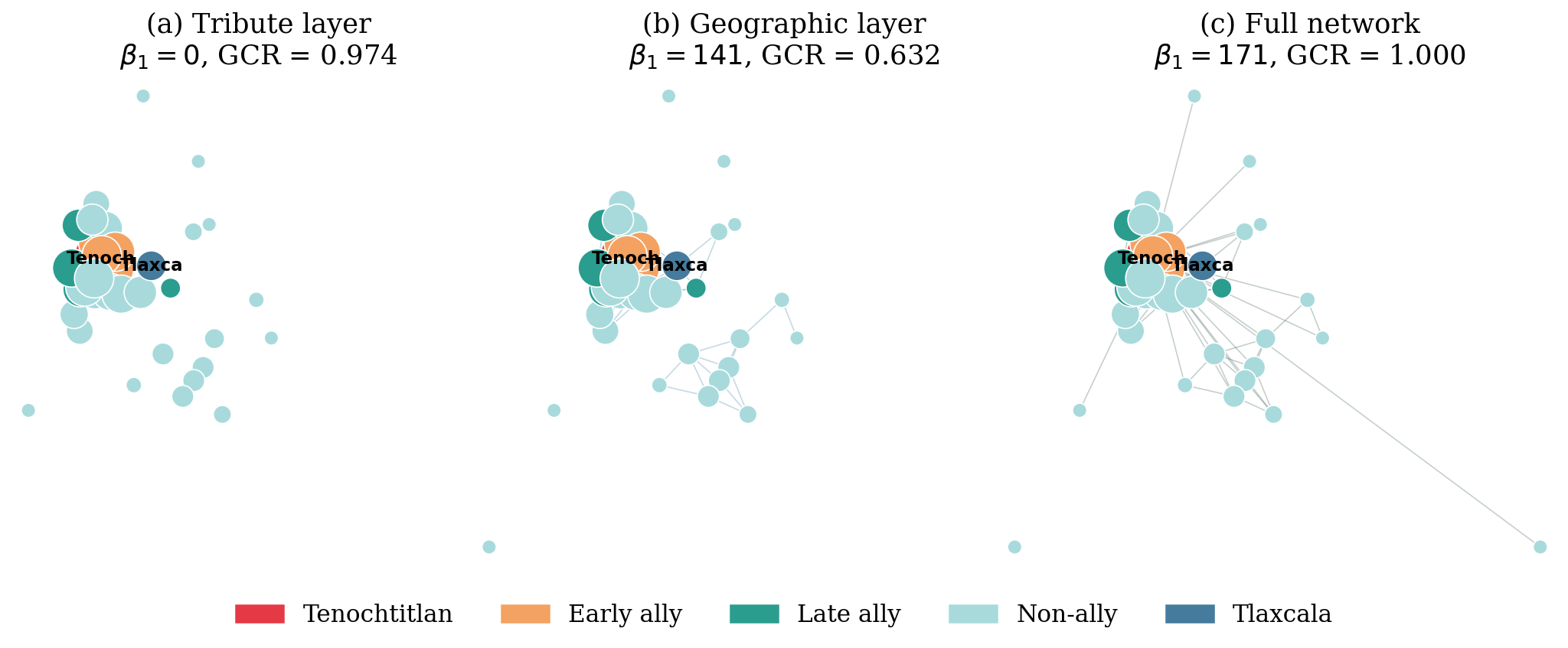}
    \caption{Multilayer network structure of the Aztec Empire.
(a) Tribute layer (36 directed edges): pure extraction tree,
$\bettione = 0$, giant component $= 97.4$\% of nodes.
(b) Geographic layer (172 undirected edges): seven components,
$\bettione = 141$, giant component $= 63.2$\%.
(c) Full network (208 edges): fully connected,
$\bettione = 171$, giant component $= 100$\%.
Node size proportional to composite centrality.
Colour codes: yellow $=$ Tenochtitlan-Centro;
red $=$ early allies (1519--1521 coalition);
orange $=$ late allies; grey $=$ non-allies.}
    \label{fig:multilayer_network}
\end{figure}
\FloatBarrier

\subsection{Robustness Under Targeted Attack}
\label{sec:robustness_results}

When the most-connected node---Tenochtitlan-Centro---is removed,
the network immediately fractures: the share of provinces that remain
mutually reachable drops from 100\% to 63\%, and 30 routing
alternatives disappear at the same stroke. This single removal,
affecting just 2.6\% of the system, produces more structural damage
than removing any twelve randomly chosen nodes. After six targeted
removals (fewer than one in six provinces),GCR reached the 0.5 value. Attacking the system by betweenness rank produces
an almost identical trajectory.
Random removal requires removing more than 12 nodes before achieving
comparable giant-component degradation.  

\begin{table}[htbp]
\caption{Robustness of the full Aztec network under targeted removal
(degree-based).  GCR: giant component ratio after $k$ removals.
$\Delta\bettione$: cumulative $\bettione$ loss.}
\label{tab:robustness}
\begin{tabular}{cccc}
\toprule
Removals ($k$) & \% nodes & GCR & $\Delta\bettione$ \\
\midrule
0 & 0.0  & 1.000 & 0  \\
1 & 2.6  & 0.630 & 30 \\
3 & 7.9  & 0.570 & 74 \\
6 & 15.8 & 0.500 & 106 \\
\midrule
\multicolumn{4}{l}{Random removal: $>12$ nodes to reach GCR $\leq 0.63$.}\\
\multicolumn{4}{l}{Asymmetry ratio: $>4$:1 (GCR), $>6$:1 ($\bettione$).}\\
\bottomrule
\end{tabular}
\end{table}

\begin{figure}[H]
    \centering
    \includegraphics[width=1\linewidth]{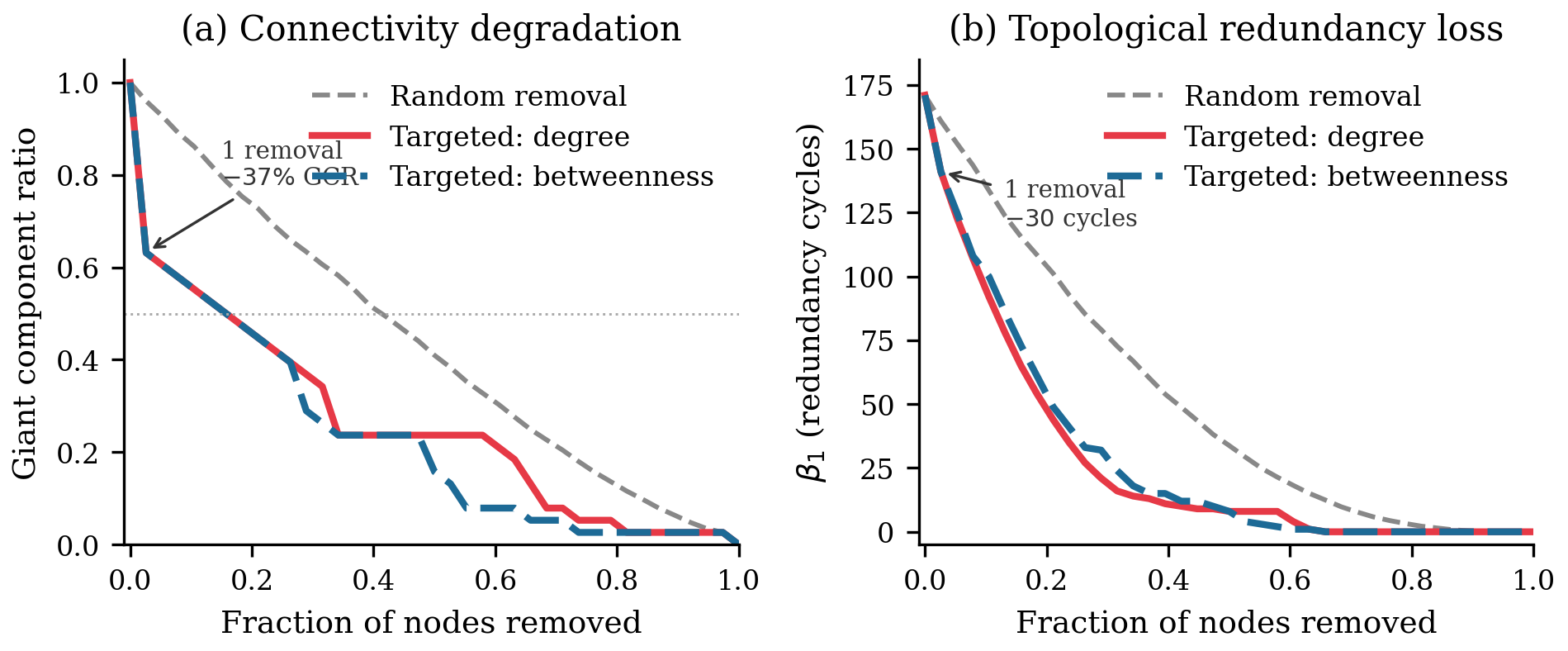}
    \caption{Robustness trajectories for the full Aztec network under
random removal (grey), targeted degree-based removal (red), and targeted
betweenness-based removal (blue).
(a) Giant component ratio as a function of fraction of nodes removed.
(b) Graph-theoretic $\bettione$ under the same strategies.
The first targeted removal (Tenochtitlan-Centro, 2.6\% of nodes)
collapses the giant component by 37\% and eliminates 30 redundancy
cycles simultaneously.  The structural asymmetry ratio $> 4$:1 is
the fragility signature.}
    \label{fig:robustness}
\end{figure}
\FloatBarrier

\subsection{Persistent Homology of the Three Layers}
\label{sec:h1results}


\begin{table}[htbp]
\caption{$\bettione$ barcode statistics by layer.
Persistence values in filtration units.
$H(\bettione)$: persistent entropy (Eq.~\ref{eq:entropy}).}
\label{tab:persistence}
\begin{tabular}{lccccc}
\toprule
Layer & $n_\mathrm{bars}$ & Max.\ pers. & Mean pers. & $H(\bettione)$ & Interp. \\
\midrule
Tribute    & 0 & ---   & ---   & 0.000 & Pure tree \\
Geographic & 2 & 0.025 & 0.020 & 0.654 & Local cycles \\
Full       & 5 & 5.041 & 1.390 & 0.797 & Radial hub \\
\bottomrule
\end{tabular}
\end{table}

The value $\bettione = 141$ reported for the geographic layer in
Table~\ref{tab:layers-tda} refers to the graph-theoretic cycle rank of the
static geographic network.  By contrast, $n_\mathrm{bars}=2$ in
Table~\ref{tab:persistence} refers to the number of persistent
$\bettione$ bars detected in the barcode under the chosen filtration.
Thus, these two quantities do not represent the same object: the former
counts independent cycles in the graph, whereas the latter counts
features that persist across the filtration.

The geographic layer produces two persistent cycles \cite{Carlsson2009}.
Some early allies, including Acolhuacan and Chalco, lie in the zone
associated with these local persistent features.  The full network
produces five cycles with a dominant class (persistence 5.04, birth 0,
death 5.04) whose generators all pass through Tenochtitlan-Centro.  The
three most persistent cycles involve Tenochtitlan paired with peripheral
nodes (Coixtlahuaca, Tepeaca, Tlapan).

\subsection{Dual Pipeline: The Tlaxcala--Puebla Void}
\label{sec:dual_results}

\begin{figure}[H]
    \centering
    \includegraphics[width=1\linewidth]{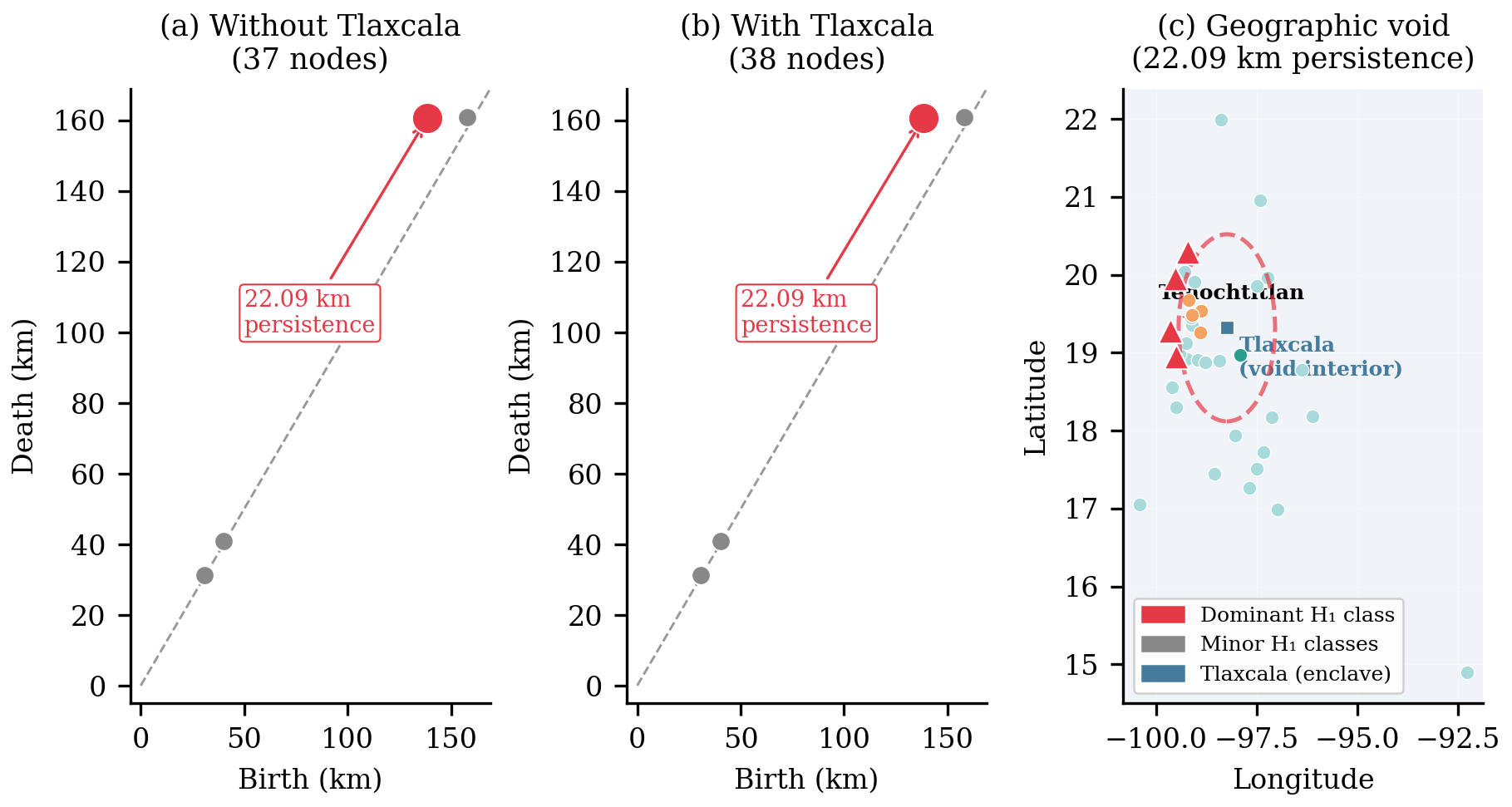}
    \caption{Dual pipeline results.
(a) $\bettione$ persistence diagram for the geographic point cloud
excluding Tlaxcala (37 nodes).  The dominant class
(persistence $= 22.09$~km, shown in red) is the Tlaxcala--Puebla void.
(b) $\bettione$ persistence diagram with Tlaxcala included (38 nodes):
as there is no changes between two cases as seen from the persistence diagrams, it confirms that Tlaxcala falls inside the
void and does not alter the cycle structure.
(c) Geographic map showing the imperial network and the ring of nodes
generating the dominant void (Malinalco, Tollocan, Xilotepec, Atotonilco
marked with triangles).  Tlaxcala shown as a star inside the ring.}
    \label{fig:dualpipeline}
\end{figure}

In the model excluding Tlaxcala (37 nodes), four $\bettione$ classes
are detected.  One dominates: born at 138.51~km, dying at
160.60~km, persistence $= 22.09$~km.  The three other classes persist
for 2.90, 0.80, and 0.52~km respectively.  The dominant class is generated by a
ring of imperial nodes (Malinalco, Tollocan, Xilotepec, Atotonilco, and
others) encircling the Tlaxcala--Puebla region at distances of 138--161~km.
The ring defines a \emph{geographic void}: a region the imperial network
surrounds but does not fill.
When Tlaxcala is added back (38 nodes), the result is exactly the same.
Tlaxcala's coordinates place it 50--95~km from its nearest imperial
neighbors (see Fig.~\ref{fig:dualpipeline}). Tlaxcala falls
\emph{inside} the void. The Tlaxcala--Puebla void is a geometric property
of the Aztec imperial domain.  It predates and is independent of the
1519 alliance.
\begin{table}[htbp]
\caption{Dual pipeline results: $\bettione$ barcode statistics for the
geographic point cloud with and without the Tlaxcala node.
Persistence values in km (Haversine distances).
The identical dominant persistence confirms that the void is a
pre-existing property of the imperial domain geometry.}
\label{tab:dual}
\begin{tabular}{lcccc}
\toprule
Model & $n_\mathrm{nodes}$ & $n_\mathrm{bars}$ & Dominant pers. (km) & Mean pers. (km) \\
\midrule
Excluding Tlaxcala & 37 & 4 & 22.09 & 6.58 \\
Including Tlaxcala & 38 & 4 & 22.09 & 6.58 \\
\midrule
\multicolumn{5}{l}{Void birth: 138.51~km; void death: 160.60~km.}\\
\multicolumn{5}{l}{All other bars: $< 3$~km (noise level).}\\
\bottomrule
\end{tabular}
\end{table}
The persistence ratio of the dominant class to the second-longest-lived
class is $22.09 / 2.90 = 7.6$. 

\FloatBarrier

\subsection{Commercial Centrality and Alliance}
\label{sec:alliance_results}


\begin{table}[htbp]
\caption{Predictor correlations with early alliance formation
(1 = early ally of the 1519--1521 coalition; $n = 37$, excluding
Tenochtitlan-Centro).}
\label{tab:alliance}
\begin{tabular}{lccc}
\toprule
Predictor & $r$ & $p$ & Interpretation \\
\midrule
Commercial centrality (geographic layer) & $+0.462$ & 0.004 & Significant \\
Composite centrality (full network)      & $+0.350$ & 0.033 & Significant \\
Distance to Tenochtitlan                 & $-0.305$ & 0.066 & Marginal \\
Tribute burden (\emph{quachtli})         & $-0.033$ & 0.845 & Non-significant \\
\midrule
Full model AUC (centrality + distance) & \multicolumn{3}{l}{0.838} \\
Commercial-centrality-only AUC         & \multicolumn{3}{l}{0.900} \\
\bottomrule
\end{tabular}
\end{table}
Tribute burden is statistically unrelated to alliance formation.
Commercial centrality in the geographic layer is the strongest single
predictor.  Early allies had, on average, twice the commercial centrality
of non-allies (0.271 versus 0.132).
  The two
redundancy cycles detected in the geographic layer
(Section~\ref{sec:h1results}) were located in the
high-commercial-centrality zone---Acolhuacan, Chalco, Tepeyacac,
Quauhtitlan---that became the early alliance core.  The mechanism of
defection was not resentment of tribute burden but access to coordination
infrastructure.

\begin{figure}[H]
    \centering
    \includegraphics[width=1\linewidth]{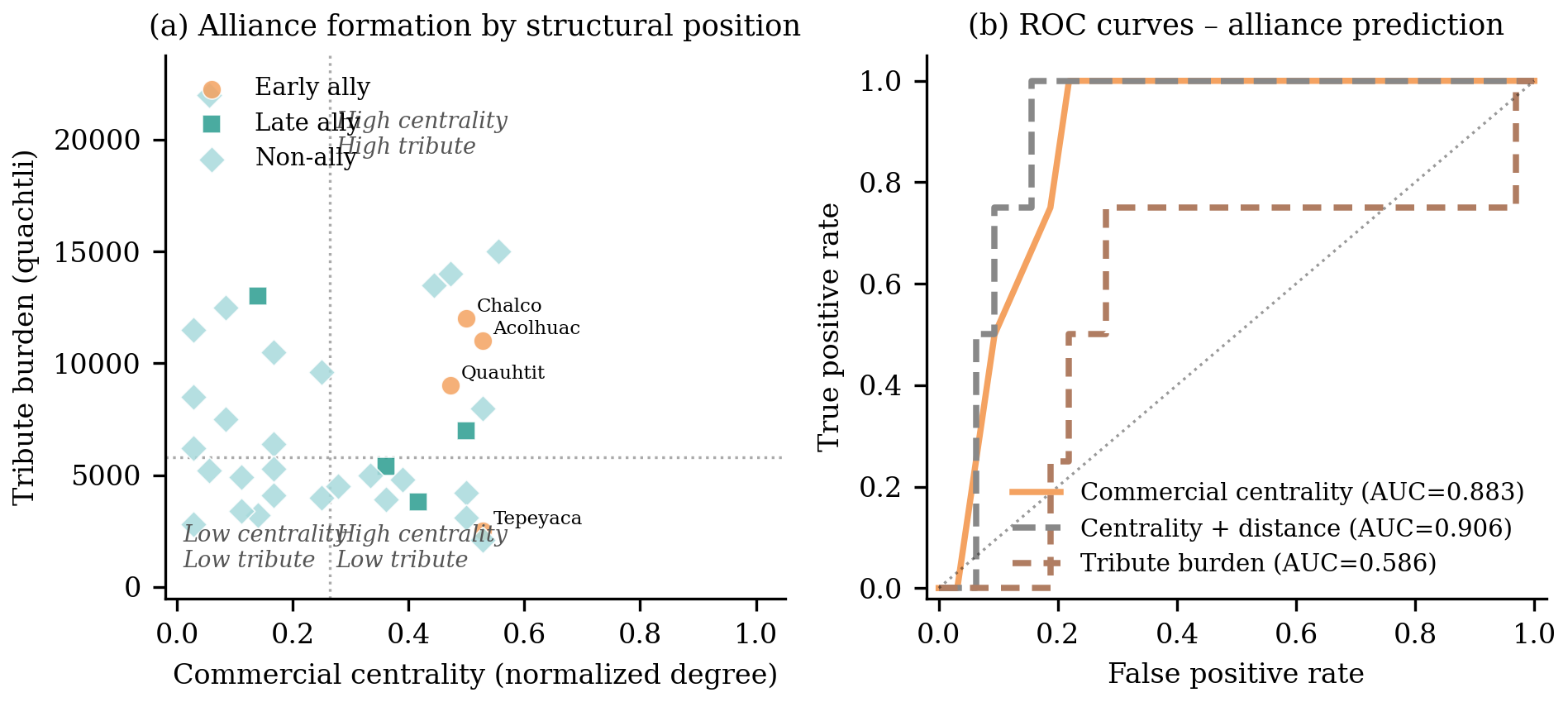}
    \caption{Alliance formation by quadrant.
Scatter plot of composite centrality versus tribute burden for all 37
provinces (excluding Tenochtitlan-Centro).
Early allies (red circles) cluster in the high-centrality,
moderate-tribute quadrant.  Late allies (orange squares) and non-allies
(grey triangles) are distributed across other quadrants.
Inset: ROC curves for the commercial-centrality model
(AUC $= 0.900$) and the baseline model (AUC $= 0.838$).}
    \label{fig:alliance}
\end{figure}
\FloatBarrier

\begin{table}[htbp]
\caption{Five principal results ($n = 38$ nodes, $n = 208$ edges).
The $\Hstar$ thresholds are $\Hstar_{\mathrm{cross}} = 0.5241$
(Roman--Byzantine calibration) \cite{BernalAlvarado2026b}.}
\label{tab:principal_results}
\begin{tabular}{clll}
\toprule
\# & Result & Metric and value & Robust \\
\midrule
1 & Tribute layer is pure tree & $\bettione = 0$, $H = 0.000$ & Y \\
2 & Hub removal collapses system & GCR: $1.000 \to 0.630$, $-30\;\bettione$ & Y \\
3 & Tlaxcala void pre-exists alliance & Dominant pers. = 22.09~km (both models) & Y \\
4 & Void 7.6$\times$ most stable feature & $22.09/2.90 = 7.6\times$ ratio & Y \\

5 & Commercial centrality drives defection & $r = +0.462$, $p = 0.004$; AUC = 0.900 & Y \\
\bottomrule
\end{tabular}
\end{table}

\section{Conclusion}
\label{sec:conclusion}

We have applied persistent homology and a dual-model topological pipeline
to a two-layer network model of the Aztec Empire (1427--1521 CE),
constructed from the \emph{Matr\'icula de Tributos} and \emph{Codex
Mendoza} records ($N = 38$ nodes, 208 edges).

The tribute layer is a pure extraction tree ($\bettione = 0$, $H = 0.000$
at all epochs), structurally more fragile than any documented point in
the Roman, Byzantine, or Han administrative datasets.  The full network's
redundancy is radially organized around Tenochtitlan-Centro, so that a
single hub removal destroys 30 redundancy cycles simultaneously and
collapses the giant component by 37\%.  

The Tlaxcala--Puebla region coincides with a $\bettione$
cycle of persistence 22.09~km in the geographic point cloud, 7.6$\times$
larger than all other cycles ($< 3$~km).  The combination of a
hub-and-spoke control structure and a dominant void adjacent to the hub
created a structural condition that facilitated the projection of force
toward the capital through the void, where no redundant routing structure
was present to slow the attack.  The two-year total collapse
(1519--1521 CE) can be interpreted as a structural consequence of the
tree-like control network.

Commercial centrality in the
geographic network ($r = +0.462$, $p = 0.004$) is the dominant predictor
of early alliance formation, not tribute burden ($r = -0.033$, $p = 0.845$).

In the Aztec system, the present model does not identify a commercial
layer institutionally independent of the extraction hierarchy. .  Instead,
the observable non-fiscal structure is the geographic layer.  The tribute
system is a pure tree ($\bettione = 0$).   What survived
the collapse was the geographic network itself---the lateral connectivity
that had enabled coalition formation and later supported the
reorganization of regional exchange under Spanish administration.

These findings place the Aztec case as a distinct mode in the
comparative typology of pre-modern imperial collapse.  The rapid
dissolution of 1519--1521 was not simply a function of Spanish military
technology, epidemic mortality, or Tlaxcalan agency, though all three
contributed; it also reflects the structural condition of an imperial
architecture with no topological redundancy in its control layer,
radially organized redundancy in its full network, and a dominant
geographic void 7.6$\times$ more persistent than any other structural
feature, occupied by the most autonomous and militarily capable enclave
in the region.

\section*{Acknowledgements}

The authors thank the institutions that make the \emph{Matr\'icula de
Tributos} and \emph{Codex Mendoza} digitizations publicly available.
We acknowledge financial support from SECIHTI and SNII (M\'exico).



\begin{thebibliography}{99}

\bibitem{BernalAlvarado2026b}
J.\ de J.\ Bernal-Alvarado, D.\ Delepine, and C.\ Pinedo Guadarrama,
Structural divergence of the Roman--Byzantine trade network, 0--1453~CE:
Persistent homology, topological velocity, and criticality indicators of
imperial collapse,arXiv:2607.05695  (2026).

\bibitem{BernalAlvarado2026c}
J.\ de J.\ Bernal-Alvarado, D.\ Delepine, and C.\ Pinedo Guadarrama,
Topological signatures of imperial stress: Persistent homology of the
Eastern Mediterranean trade network, 0--400~CE, arXiv:2605.27200  (2026).

\bibitem{BernalAlvarado2026d}
J.\ de J.\ Bernal-Alvarado, D.\ Delepine, and C.\ Pinedo Guadarrama,
Topological signatures of imperial collapse and fragmentation:
Administrative dissolution, territorial reorganization, and early-warning
observables in the Han dynasty network (206~BCE--220~CE), arXiv:2607.09010
(2026).


\bibitem{Albert2000}
R.\ Albert, H.\ Jeong, and A.-L.\ Barab\'asi,
Error and attack tolerance of complex networks,
\emph{Nature} \textbf{406}, 378 (2000).

\bibitem{Berdan1992}
F.\ F.\ Berdan,
\emph{The Aztecs of Central Mexico: An Imperial Society}
(Harcourt Brace, Fort Worth, 1992).

\bibitem{BerlinBerdan1996}
F.\ F.\ Berdan and P.\ R.\ Anawalt (Eds.),
\emph{The Essential Codex Mendoza}
(University of California Press, Berkeley, 1996).

\bibitem{Carlsson2009}
G.\ Carlsson,
Topology and data,
\emph{Bull.\ Am.\ Math.\ Soc.}\ \textbf{46}, 255 (2009).

\bibitem{Davies1973}
N.\ Davies,
\emph{The Aztecs: A History}
(Macmillan, London, 1973).

\bibitem{Edelsbrunner2002}
H.\ Edelsbrunner, D.\ Letscher, and A.\ Zomorodian,
Topological persistence and simplification,
\emph{Discrete Comput.\ Geom.}\ \textbf{28}, 511 (2002).

\bibitem{GUDHI2024}
The GUDHI Project,
GUDHI user and reference manual, version~3.12.0
(GUDHI Editorial Board, 2024).

\bibitem{Hassig1985}
R.\ Hassig,
\emph{Trade, Tribute, and Transportation: The Sixteenth-Century Political
Economy of the Valley of Mexico}
(University of Oklahoma Press, Norman, 1985).

\bibitem{Hassig1994}
R.\ Hassig,
\emph{Mexico and the Spanish Conquest}
(Longman, London, 1994).


\bibitem{Lockhart1993}
J.\ Lockhart,
\emph{The Nahuas After the Conquest}
(Stanford University Press, Stanford, 1993).

\bibitem{McCormick2001}
M.\ McCormick,
\emph{Origins of the European Economy: Communications and Commerce,
A.D.\ 300--900}
(Cambridge University Press, Cambridge, 2001).

\bibitem{Restall2003}
M.\ Restall,
\emph{Seven Myths of the Spanish Conquest}
(Oxford University Press, Oxford, 2003).

\bibitem{Rucco2016}
M.\ Rucco, F.\ Castiglione, E.\ Merelli, and M.\ Pettini,
Characterisation of the idiotypic immune network through persistent
entropy,
\emph{Entropy} \textbf{18}, 117 (2016).

\bibitem{Smith1986}
M.\ E.\ Smith,
The role of social stratification in the Aztec Empire: A view from the
provinces,
\emph{Am.\ Anthropol.}\ \textbf{88}, 70 (1986).

\bibitem{Thomas1993}
H.\ Thomas,
\emph{Conquest: Moctezuma, Cort\'es, and the Fall of Old Mexico}
(Simon \& Schuster, New York, 1993).

\bibitem{Townsend2019}
C.\ Townsend,
\emph{Fifth Sun: A New History of the Aztecs}
(Oxford University Press, Oxford, 2019).

\bibitem{Zomorodian2005}
A.\ Zomorodian and G.\ Carlsson,
Computing persistent homology,
\emph{Discrete Comput.\ Geom.}\ \textbf{33}, 249 (2005).

\bibitem{Pearson1895}
K.~Pearson, ``Notes on regression and inheritance in the case of two parents,''
Proceedings of the Royal Society of London \textbf{58}, 240--242 (1895).

\bibitem{Hanley1982}
J.~A.~Hanley and B.~J.~McNeil, ``The meaning and use of the area under
a receiver operating characteristic (ROC) curve,''
Radiology \textbf{143}, 29--36 (1982).

\bibitem{Fawcett2006}
T.~Fawcett, ``An introduction to ROC analysis,''
Pattern Recognition Letters \textbf{27}, 861--874 (2006).

\bibitem{Cohen2000}
R.~Cohen, K.~Erez, D.~ben-Avraham, and S.~Havlin,
``Resilience of the Internet to random breakdowns,''
Phys.\ Rev.\ Lett.\ \textbf{85}, 4626 (2000).

\bibitem{Callaway2000}
D.~S.~Callaway, M.~E.~J.~Newman, S.~H.~Strogatz, and D.~J.~Watts,
``Network robustness and fragility: Percolation on random graphs,''
Phys.\ Rev.\ Lett.\ \textbf{85}, 5468 (2000).

\bibitem{Holme2002}
P.~Holme, B.~J.~Kim, C.~N.~Yoon, and S.~K.~Han,
``Attack vulnerability of complex networks,''
Phys.\ Rev.\ E \textbf{65}, 056109 (2002).

\bibitem{Hosmer2013}
D.~W.~Hosmer, S.~Lemeshow, and R.~X.~Sturdivant,
\emph{Applied Logistic Regression}, 3rd ed.
(Wiley, Hoboken, 2013).

\end{thebibliography}
\end{document}